\documentstyle[12pt,fleqn,amssymb]{article}
\newcommand{\< }{$}
\renewcommand{\>}{$}
\newcommand {\bR}{{\Bbb R}}
\newcommand {\bN}{{\Bbb N}}
\newcommand {\bP}{{\Bbb P}}
\newcommand {\bZ}{{\Bbb Z}}
\newcommand {\bC}{{\Bbb C}}
\newtheorem{theorem}{Theorem} 
\newtheorem {lemma}[theorem]{Lemma}
\newtheorem {propo}[theorem]{Proposition}

\newtheorem {coro}[theorem]{Corollary}

\renewcommand {\l}{\left}
\newcommand {\ri}{\right}
\newcommand {\vep}{\varepsilon}
\newcommand {\ar}{\rightarrow}

\newcommand {\eh}{\textstyle \frac{1}{2}}
\newcommand {\Euler}{\varphi}
\renewcommand {\Re}{{\rm Re}} 
\newcommand {\hcb}{{\bf h}^{C}}   
\newcommand {\hgb}{{\bf h}^{G}}   
\newcommand {\Hcb}{{\bf H}^{C}}   
\newcommand {\Hgb}{{\bf H}^{G}}   
\newcommand {\Gb}{{\bf G}}    
\newcommand {\jg}{j^{G}}      
\newcommand {\Fou}{{\cal F}}  

\newcommand {\LA}{\left\langle}
\newcommand {\RA}{\right\rangle}
\newcommand {\cE}{{\cal E}}
\newcommand{\idty}{{\rm 1\mskip-4mu l}} 
\newcommand{\beq}{\begin{equation}}
\newcommand{\eeq}{\end{equation}}
\newcommand{\beqn}{\begin{eqnarray}}
\newcommand{\eeqn}{\end{eqnarray}}
\newcommand{\beqno}{\begin{eqnarray*}}
\newcommand{\eeqno}{\end{eqnarray*}}
\newcommand{\Leq}[1]{\label{#1}\end{equation}}		   
\newcommand{\bem}{\l(\! \begin{array}}
\newcommand{\eem}{\end{array}\!\ri)}
\newcommand{\NN}{\nonumber}
\newcommand {\qmbox}[1]{\quad\mbox{#1}\quad}
\newcommand {\cC}{{\cal C}}
\newcommand {\cS}{{\cal S}}

\newcommand {\nmax}{{n_{\rm max}}}

\begin{document}
\title {A Fully Magnetizing Phase Transition}
\date{\today}
\author{Pierluigi Contucci\thanks{Department of Mathematics, University 
of California, Davis CA 95616-8633, USA. email: contucci@math.ucdavis.edu}
\and 
Peter Kleban\thanks{LASST and Department of Physics and Astronomy,
University of Maine, Orono, ME 04469, USA. e-mail: kleban@maine.edu}
\and
Andreas Knauf\thanks{Max-Planck-Institute for Mathematics in the Sciences, 
Inselstr.\ 22--26, D-04103 Leipzig, Germany. 
e-mail: knauf@mis.mpg.de
}}
\maketitle
\begin{abstract}
We analyze the Farey spin chain, a one dimensional spin system
with effective interaction decaying like the squared inverse distance. 
Using a polymer model technique, we show that 
when the temperature is decreased below the (single) critical temperature
\<T_c=\eh\> , the magnetization jumps from zero to one.
\end{abstract}
\section{Introduction}
Can a magnet keep its full mean magnetization \< \LA m\RA=1 \> 
up to the Curie temperature \< T_c \> and then loose it at one stroke?
Definitely such a property would be different from the usual situation,
where \< \LA m\RA \> continuously decreases to zero 
(though not being differentiable at \>T_c \>), or jumps discontinuously
by an amount strictly less than the saturation value.

It has been proven \cite{ACCN,AN,Dy1,Dy2,Ru,Th} 
that certain spin chains of long range ferromagnetic 
interaction exhibit a discontinuity of \< \LA m\RA \> 
at \< T_c \>, jumping from a value in the interval \< (0,1) \> to zero.

In one dimension such a phenomenon can only occur if the effective interaction
between spins of distance \< d \> decays at most like \< d^{-2} \>,
since there cannot be a phase transition for a decay rate of \< d^{-\alpha} \>
if \< \alpha>2 \>.

However these examples do not exactly provide a positive answer to the question
posed initially, since the jump of \< \LA m\RA \> 
at \< T_c \> is strictly smaller than one.
Indeed for non-zero temperatures a mean magnetization \< \LA m\RA=\pm 1 \> 
is only possible if the forces between the spins become so strong that
one may legitimately ask whether this could endanger the 
existence of a thermodynamic limit.

In the present paper we show the contrary by considering the example of the 
Farey fraction spin chain. Similar to  \cite{GK} the abstract polymer model
formalism, is introduced in {\em Sect.\ 2}. Since the limit free 
energy coincides with the one of the number-theoretical spin chain 
({\em Theorem \ref{fe}}\,), the single phase transition 
(nonanalyticity of the free energy density) is situated at 
inverse temperature \< \beta=2 \>.

In {\em Sections 5 {\em resp.}\,6} we consider the mean square magnetization 
in the regimes below resp.\ above the temperature. Whereas
\<\LA m^2\RA(\beta) =1\> for low temperatures ({\em Theorem \ref{low}}\,),
this quantity vanishes above \< T_c\>  ({\em Theorem \ref{high}}\,).

We conjecture, and plan to prove, that the spin chain 
has exactly two extremal Gibbs measures in its
low temperature phase, and one above \<T_c\> .

We also invoke a polymer model technique
similar to the one developed in \cite{GK} to estimate the
strength of the interaction.
\section{The Model}
In \cite{Kn1} the so-called {\em number-theoretical spin chain} was introduced,
whose low-temperature partition function equals a quotient of 
Riemann zeta functions. In a series of subsequent papers (see \cite{Knx} for a
survey) this model was then analyzed further. In particular it was shown in
\cite{CK} that a phase transition with a jump of \< m \> from one to zero occurs at
 \>T_c=\eh \>.

The number-theoretical spin chain shows an asymptotic decay of interactions
which is exactly of the form \< d^{-2} \>, and the limit free energy density exists.
The main motivation of its study lies in its connection with number theory, 
and more specifically in the hope that its ferromagnetic character together 
with a version of the Lee-Yang theorem could shed a light on the location of the
zeroes of the Riemann zeta function.

>From the statistical mechanics point of view it should, however, be said that it
lacks the strict symmetries usually encountered in ferromagnets. It is neither
fully translation invariant nor invariant under spin reversal, although
both symmetries are asymptotically present in the bulk.

In \cite{KO} the so-called {\em Farey fraction spin chain}
was introduced as a spin system of statistical mechanics related to the
Farey fractions in number theory.
As we shall state below, this chain, which has strong relations with the one
mentioned above, but a less direct 
number-theoretical interpretation of its partition function,
has all relevant symmetries. 

The definition of the Farey chain in \cite{KO} was based on
functions
\[M_k:\Gb_k\ar{\rm SL}(2,\bZ)\qquad(k\geq 0)\]
on the additive group \< \Gb_k:=\{0,1\}^{\{1,\ldots,k\}} \>, 
inductively defined by setting \< \textstyle M_0:={1\ 0\choose 0\ 1} \>
and for \< k\geq1 \>
\beq
M_{k}(\sigma):=
A^{1-\sigma_{k}}B^{\sigma_{k}} M_{k-1}(\sigma_1,\ldots,\sigma_{k-1})
\qquad(\sigma\in\Gb_k),
\Leq{def:Mk}
with \< \textstyle A:={1\ 0\choose 1\ 1} \> and 
 \>\textstyle B:= A^t = {1\ 1\choose 0\ 1} \>.
The function
\[E_k:=\ln(T_k) \qmbox{with} T_k:={\rm Trace}(M_k):\Gb_k\ar\bN\]
was interpreted as the {\em energy function} of a spin chain with \< k \> spins
with values \< \sigma_1,\ldots,\sigma_k \>.

Then by discrete Fourier transformation
\beq
(\Fou_k f)(t) := 
2^{-k} \sum_{\sigma\in\Gb_k} f(\sigma) \cdot (-1)^{\sigma\cdot t}
\qquad (t\in\Gb_k)
\Leq{Fourier}
the energy function has the form
\[E_k(\sigma)= -\sum_{t\in\Gb_k} J_k(t) (-1)^{\sigma\cdot t}
\qquad(\sigma\in\Gb_k)\]
with the so-called {\em interaction coefficients}
\[J_k(t) := -(\Fou_k E_k)(t) \qquad(t\in\Gb_k).\]
The `lattice gas' spin values \< \sigma_i=0,1 \> are used here
for convenience. The mean magnetization
\[ m_k:=\frac{1}{k}\sum_{i=1}^k s_i, \]
however, is defined using the spin values
 \>s_i(\sigma):=(-1)^{\sigma_i}\in\{\pm1\} \>.

The Farey spin chain has the following symmetries:
\begin{enumerate}
\item
When one interprets \< \{1,\ldots,k\} \> as a system of representatives
of the residue class ring \< \bZ/k\bZ=\{l+k\bZ\mid l\in\bZ\} \>, then 
by cyclicity of the trace the energy function is invariant under the shift
\beq
\cS_k:\Gb_k\ar\Gb_k\qmbox{,}\cS_k(\sigma)_l:=\sigma_{l-1}
\Leq{shift}
on the configuration space \< \Gb_k \> of the chain.
So the interaction is translation-invariant, too ( \>J_k\circ\cS_k=J_k \>).
\item
Since \< AP=PB \> for \< \textstyle P:={0\ 1\choose 1\ 0} \>
\[M_k(\sigma_k,\ldots,\sigma_{1})=PM_k(\sigma_1,\ldots,\sigma_{k})^tP.\]
This implies the mirror symmetry
\[E_k(\sigma_k,\ldots,\sigma_{1})=E_k(\sigma_1,\ldots,\sigma_{k})\]
and a similar relation for the interaction coefficients.
\item
Finally we notice that by 2) and transposition invariance of the trace
\[E_k(1-\sigma)=E_k(\sigma)\qmbox{for} 
1-\sigma:=(1-\sigma_1,\ldots,1-\sigma_{k})\]
so that 
\[J_k(t)=0 \qmbox{for}\sum_{i=1}^k t_i\mbox{ odd}. \]
\end{enumerate}
By 3) we need only consider \< t \> in the {\em even} subgroup 
\[\Gb_k^e:=\textstyle \l\{t\in\Gb_k\mid\sum_{i=1}^k t_i\mbox{ even}\ri\}. \]

 \>J_k(0)<0 \>, since this is the negative mean of the 
(positive) energy function \< E_k \>. Note that this is the only interaction
coefficient which does not influence the Gibbs measure.

\section{A Polymer Model Interpretation}
The notion of polymer models grew out of an abstraction of situations like the
one encountered in the low temperature expansion of the Ising model.
There one may decompose contours \< X \> 
into non-intersecting cycles \< \gamma_i \> ( \>X=(\gamma_1,\ldots,\gamma_l) \>), 
and express the 
Boltzmann factor of the spin configuration in terms of products of
activities  attributed to these cycles 
(the activity \< z(\gamma_i) \> of a cycle equals the exponential
of its length, multiplied with minus the inverse temperature).

In an abstract setting (see, e.g., 
Gallavotti, Martin-L\"{o}f and Miracle-Sol\'{e} \cite{GMM},
Glimm and Jaffe \cite{GJ} and Simon \cite{Si}) 
one starts with a set \< P \> (which we assume here to be finite),
whose elements are called {\em polymers}. 
Two given polymers \< \gamma_1,\gamma_2\in P \> may or may not overlap
(be {\em incompatible}). Incompatibility is assumed to be a
reflexive and symmetric relation on \< P \>.

Thus one may associate to a \< l \>-{\em polymer}
 \>X:=(\gamma_1,\ldots,\gamma_l)\in P^l \> an undirected graph 
 \>G(X)=(V(X),E(X)) \> 
with vertex set \< V(X):=\{1,\ldots,l\} \>, vertices \< i\neq
j \> being connected by the edge \< \{i,j\}\in E(X) \> if 
 \>\gamma_i \> and \< \gamma_j \> are incompatible.
Accordingly the \< l \>-polymer \< X \> is called {\em connected} if \< G(X) \> is
path-connected and {\em disconnected} if it has no edges ( \>E(X)=\emptyset \>).

The corresponding subsets of \< P^l \> are called \< C^l \> resp.\ \< D^l \>,
with \< D^0:=P^0:=\{\emptyset\} \> consisting of a single element. Moreover
 \>P^{\infty}:= \bigcup_{l=0}^{\infty} P^l \> with the subsets
 \>D^{\infty}:= \bigcup_{l=0}^{\infty} D^l \> and 
 \>C^{\infty}:= \bigcup_{l=1}^{\infty} C^l \>.
We write \< |X|:=l \> if \< X\in P^l \>.

Statistical weights or {\em activities} \< z:P\ar\bC \> of the
polymers are multiplied to give the activities 
 \>z^X:=\prod_{i=1}^l z(\gamma_i) \> of \< l \>-polymers \< X \>.
A system of statistical mechanics is called {\em polymer model} if
its partition function \< Z \> has the form
\beq
Z = \sum_{X\in D^{\infty}_\Lambda} \frac{z^X}{|X|!}.
\Leq{eq:parti}
Then, up a normalization factor, the free energy is given by
\beq
\ln(Z) = \sum_{X\in C^{\infty}_\Lambda} \frac{n(X)}{|X|!}z^X,
\Leq{eq:freie}
with \< n(X):=n_+(X)-n_-(X) \>, \< n_\pm (X) \> being the number of
subgraphs of \< G(X) \> connecting the vertices of \< G(X) \> 
with an even resp.\ odd number of edges (see Gallavotti et al.\
\cite{GMM}). It is known that 
(see, e.g. Prop.\ 20.3.5 of \cite{GJ}) that
\beq
(-1)^{|V|-1}n(G)\geq 0.
\Leq{sign:n}
This also follows from the {\em deletion-contraction property}
\beq
n(G)=n(G')-n(G''),
\Leq{del:con}
where \< G' \> is obtained from \< G \> by deleting an edge and \< G'' \> 
is the graph which arises by contracting the same edge of \< G \> (see, e.g.\ 
Read \cite{Re}).

In the present context of a chain with \< k \> spins we use 
\begin{itemize}
\item
the set \< P_k \>
of \< 1+k(k-1) \> polymers given by
\[P_k:=\{p\}\cup \{p_{l,r}\mid l\neq r\in\bZ/k\bZ\}.\] 
\item
We map the polymers \< \gamma\in P_k \> to group elements
\< \hat{\gamma}\in\Gb_k^e \> by setting \< \hat{p}:=0 \> and 
\<\hat{p}_{l,r}:=\delta_{l}+\delta_{r},\>
where for \< i\in\bZ/k\bZ \> the group
element \< \delta_i \> has the form \< \delta_i(l)=1 \> if \< l=i \> and zero otherwise.

This map induces a map 
\[X=(\gamma_1,\ldots,\gamma_l)\mapsto \hat{X}:=\sum_{i=1}^l \hat{\gamma}_i\]
from the set \< P_k^{\infty} \> of multi-polymers to \< \Gb_k^e \>.
\item
The {\em support} of our polymers is given by \< {\rm supp}(p):=\bZ/k\bZ \> and
\[{\rm supp}(p_{l,r}) := \{l,l+1,\ldots,r-1,r\}\subset\bZ/k\bZ.\]
Note that the polymer \< p_{r,l} \> is different from 
 \>p_{l,r} \>, although the group elements \< \hat{p}_{l,r} \> and \< \hat{p}_{r,l} \>
coincide, and for \< k=2 \> the supports 
 \>{\rm supp}(p)={\rm supp}(p_{1,2})={\rm supp}(p_{2,1}) \>.

The polymers \< \gamma \> and \< \gamma' \> are called overlapping or {\em
incompatible} if
\[{\rm supp}(\gamma)\cap{\rm supp}(\gamma')\neq \emptyset.\]
\item
We attribute to the polymers the activities 
\beq
z(p) := 3^{-|{\rm supp}(p)|}=3^{-k}
\qmbox{and}z(p_{l,r}) := -3^{-|{\rm supp}(p_{l,r})|}.
\Leq{act}
\end{itemize}
Every group element \< t\in\Gb^e_k \> allows for exactly two representations
 \>t=\hat{X} \> by disjoint multi-polymers \< X\in D_k^{\infty} \>.

\begin{lemma} \label{l1}
The Fourier transform \< j_k:=\Fou_k T_k \> can be written as
\beq
j_k(t) =
\l(\textstyle \frac{3}{2}\ri)^k \sum_{X\in D_k^{\infty}:\hat{X}=t} z(X) 
\qquad(t\in\Gb_k).
\Leq{poly}
\end{lemma}
{\bf Proof.} 
\begin{itemize}
\item
For \< t \> odd both sides are zero.
\item
For \< t=0 \> we perform the sum to obtain
\[j_k(0)=2^{-k}\sum_{\sigma\in\Gb_k} {\rm Trace}(M_k(\sigma))
=2^{-k}{\rm Trace}(S^k)\]
with \< \textstyle S:=A+B={2\ 1\choose 1\ 2} \>, which has eigenvalues one 
and three.
So \< j_k(0)=(3^k+1)/2^k \>. On the other hand, the r.h.s\ of (\ref{poly})
is of the form 
\[\l(\textstyle \frac{3}{2}\ri)^k (z(\emptyset)+z(p))=
\l(\textstyle \frac{3}{2}\ri)^k (1+3^{-k}).\]
\item
For \< t\in \Gb_k^e\setminus\{0\} \> we assume without loss of generality, 
using cyclicity of the trace, that 
\[t=(0_{m_1},1,0_{m_2},1,\ldots,1,0_{m_{2n}},1)\]
with \< 0_m=(0,\ldots,0)\in\Gb_m \>.
Then with \< \textstyle D:=A-B={\,0\ -1\choose 1\ \ 0} \> 
\[j_k(t) = 2^{-k} {\rm Trace}\l(S^{m_1}DS^{m_2}D\ldots DS^{m_{2n}}D\ri).\]
Now 
\beq
DS^{m}D=S^m-(3^m+1)\idty
\Leq{dsd}
commutes with \< S \>, and \< D^2=-\idty \> so that
\[j_k(t) = (-1)^{n-1}2^{-k} {\rm Trace}\l(S^{\Delta m_1}DS^{\Delta m_2}D\ri)\]
with \< \Delta m_1:=\sum_{l=1}^n m_{2l-1} \> and 
 \>\Delta m_2:=\sum_{l=1}^n m_{2l} \>.
Thus using (\ref{dsd}) we arrive at 
\beqno
j_k(t)&=& (-1)^{n-1}2^{-k} 
{\rm Trace}\l(S^{\Delta m_1+\Delta m_2}-(1+3^{\Delta m_2})S^{\Delta m_1}\ri)\\
&=& (-1)^{n}2^{-k} \l(-(3^{\Delta m_1+\Delta m_2}+1)+
(3^{\Delta m_1}+1)(3^{\Delta m_2}+1)\ri)\\
&=& (-1)^{n}2^{-k}\l(3^{\Delta m_1} + 3^{\Delta m_2}\ri).
\eeqno
\end{itemize}
This equals the r.h.s.\ of (\ref{poly}).
\hfill \>\Box \>
\section{Comparison with the Number-Theoretical Spin Chain}
The number-theoretical spin chain of length \< k \> has the 
{\em canonical} energy function
\[\Hcb_k := \ln(\hcb_k)\qmbox{with}\hcb_k:\Gb_k\ar\bN\] 
inductively defined by
\beq
\hcb_0 := 1,\quad 
\hcb_{k+1}(\sigma,\sigma_{k+1}) :=
\hcb_{k}(\sigma)+\sigma_{k+1}\hcb_{k}(1-\sigma),\qquad(\sigma\in\Gb_k).
\Leq{def:hck}
It turns out to be useful to consider the {\em grand canonical} energy functions
\[ \Hgb_k:\Gb_k\ar\bR\qmbox{with}\Hgb_k(\sigma):=\Hcb_{k+1}(\sigma,1)\qquad
(\sigma\in\Gb_k),\]
too, which is the logarithm of 
\beq
\hgb_k:\Gb_k\ar\bN\qmbox{,}
\hgb_{k}(\sigma):=\hcb_{k+1}(\sigma,1)=\hcb_{k}(\sigma)+\hcb_{k}(1-\sigma).
\Leq{def:hg}
Namely in \cite{GK} polymer model techniques
were applied to estimate the grand canonical interaction. 
 \> \jg_k := -\Fou_k \Hgb_k  \>. These were applied to the subset
\beq
\tilde{P}_k := \{ p_{l,r}\in P_k \mid l<r \}
\Leq{tilde:P}
of polymers (where the inequality \<  <  \> in 
\<  \bZ/k\bZ  \> is understood as the one
for the representatives in \<  \{1,\ldots,k\} \>). 
This is the set of polymers which contribute to the thermodynamic limit.

For \< t\in\Gb_k\setminus\{0\} \> the resulting formula 
\beqn
\jg_k(t) = - \delta_{t,0}\cdot (\ln(2) + k\ln(3/2)) - 
\sum_{\stackrel{X\in \tilde{C}^{\infty}_k}{\hat{X}=t}} 
\frac{n(X)}{|X|!} z^X 
\label{GGG}
\eeqn
for these grand canonical
interaction coefficients contains only nonnegative terms.
This follows from (\ref{sign:n}) and the fact that all activities (\ref{act})
of polymers in \< \tilde{P}_k \> are negative. 
Similarly the canonical interaction of the number-theoretical spin chain was
shown to be ferromagnetic.
\begin{lemma} \label{l2}
The Farey interaction coefficients \< J_k(t)=-\Fou_k E_k(t) \> can be written as
\beq
J_k(t) =- \delta_{t,0}k\ln(3/2) - 
\sum_{\stackrel{X\in {C}^{\infty}_k}{\hat{X}=t}} 
\frac{n(X)}{|X|!} z^X 
\qquad(t\in\Gb_k).
\Leq{F:poly}
\end{lemma}
{\bf Proof.}
Since \< j_k:=\Fou_k T_k \>, we have \< E_k=2^{k}\Fou_k j_k \> and
\beqn
J_k(t) &=& - 2^{-k} \sum_{\sigma\in\Gb_k} E_k(\sigma) \cdot(-1)^{\sigma\cdot t} \NN \\
&=& - 2^{-k} \sum_{\sigma\in\Gb_k} 
\ln\l[\sum_{s\in\Gb_k} j_k(s) \cdot(-1)^{s\cdot \sigma}\ri]
\cdot(-1)^{\sigma\cdot t} \NN \\
&=& - \delta_{t,0}\cdot k\ln(3/2) - 
2^{-k} \sum_{\sigma\in\Gb_k} \ln\l[\sum_{X\in D_k^{\infty}}
\frac{\tilde{z}_\sigma^X}{|X|!}\ri]
\cdot(-1)^{\sigma\cdot t}
\label{CC}
\eeqn
where the redefined single-polymer activities 
 \>\tilde{z}_\sigma(\gamma) \>, \< \gamma\in P_k \> are given by 
\[\tilde{z}_\sigma(\gamma) := 
z_\sigma(\gamma)\cdot (-1)^{\sigma\cdot\hat{\gamma}},\] 
that is \< \tilde{z}_\sigma(p) = z(p) \> and 
 \>\tilde{z}_\sigma(p_{l,r}) = z(p_{l,r})\cdot (-1)^{\sigma_l + \sigma_r} \>.
By (\ref{eq:freie}) we get 
\beqn
\lefteqn{J_k(t) + \delta_{t,0}\cdot k\ln(3/2)} \NN \\
&=& - 2^{-k} \sum_{\sigma\in\Gb_k} \sum_{X\in C^\infty_k} 
\frac{n(X)}{|X|!} \tilde{z}_\sigma^X\cdot  (-1)^{\sigma\cdot t} \NN  \\
&=& - \sum_{X\in C^\infty_k} \frac{n(X)}{|X|!} z^X \cdot  
2^{-k} \sum_{\sigma\in\Gb_k} (-1)^{\sigma\cdot (t+\hat{X})} 
= - \sum_{\stackrel{X\in C^\infty_k}
           {\hat{X}=t}} \frac{n(X)}{|X|!} z^X,
\label{CCC}
\eeqn
using the identity \< \sum_{\sigma\in\Gb_k} (-1)^{\sigma\cdot s} = 
2^k \delta_{s,0} \>.\hfill \>\Box \>\\[2mm]
Although formula (\ref{F:poly}) looks very similar to (\ref{GGG}),
the sum is over all connected multipolymers based on the full set \< P_k \>
of polymers, instead of the subset (\ref{tilde:P}). 
Therefore not all terms in that sum are positive. 
By (\ref{sign:n}) and (\ref{act}) the negative terms 
are precisely the ones containing an odd number of copies 
of the polymer \< p \>. 
Thus the positivity of the interaction for {\em finite} \< k\> 
does not follow immediately.
\section{The Free Energy}
\begin{theorem} \label{fe}
The limit free energy density
\[F(\beta):=\lim_{k\ar\infty} F_k(\beta)\qmbox{of}
F_k(\beta):=\frac{-1}{k\beta}\ln(Z_k(\beta)) \qquad(\beta>0)\]
with \< k \>-spin partition function 
 \>Z_k(\beta):=\sum_{\sigma\in\Gb_k} \exp(-\beta E_k(\sigma)) \> 
exists and equals the one of the number-theoretical spin chain.
\end{theorem}
{\bf Proof.} 
We use the canonical and grand canonical ensembles as bounds for \< F_k \>.
{\bf 1)} 
Since the entries of the matrices \< M_k(\sigma) \> are non-negative,
an upper bound for \< T_k={\rm Trace}(M_k) \> is given by 
 \>{\rm Trace}\l({1\ 1\choose 1\ 1} M_k\ri) \>.
But for \< \sigma\in\Gb_k \>
\beq
\hspace{-5mm}
\textstyle  {\rm Trace}\l({1\ 0\choose 1\ 0}  M_k(\sigma)\ri) = 
\hcb_k(\sigma) \qmbox{and}
{\rm Trace}\l({0\ 1\choose 0\ 1}  M_k(\sigma)\ri) = \hcb_k(1-\sigma), 
\Leq{relation}
since both sides equal one for \< k=0 \>, and
\beqno
\lefteqn{\textstyle  {\rm Trace}\l({1\ 0\choose 1\ 0}  M_{k}(\sigma)\ri)}\\ 
&=& 
\textstyle {\rm Trace}\l({1\ 0\choose 1\ 0}  
A^{1-\sigma_k}B^{\sigma_k} 
M_{k-1}(\sigma_1,\ldots,\sigma_{k-1})
\ri)\\
&=& \textstyle
{\rm Trace}\l(\l({1\ 0\choose 1\ 0}+\sigma_k {0\ 1\choose 0\ 1}\ri) 
 M_{k-1}(\sigma_1,\ldots,\sigma_{k-1})\ri)\\
 &=&\textstyle\hcb_{k-1}(\sigma_1,\ldots,\sigma_{k-1})+
\sigma_k\hcb_{k-1}(1-\sigma_1,\ldots,1-\sigma_{k-1})
=\hcb_k(\sigma)
\eeqno
and similar for the second identity in (\ref{relation}).
Adding these identities and, using Def.\ (\ref{def:hg}), shows that
\[T_k\leq \hgb_k\qquad(k\in\bN_0).\]
{\bf 2)}
To derive a lower bound for \< T_k \>, we notice that for \<  \sigma\in\Gb_{k-1}  \>
$$\textstyle  {\rm Trace}\l({1\ 0\choose 0\ 0}  M_k(0,\sigma)\ri) = 
\hcb_{k-1}(\sigma)\mbox{ and }
{\rm Trace}\l({0\ 1\choose 0\ 0}  M_k(0,\sigma)\ri) = 
\hcb_{k-1}(1-\sigma)
,$$
since both sides equal one for \< k=1 \>, and for \< \sigma\in\Gb_{k-1} \>
\beqno
\lefteqn{\textstyle  {\rm Trace}\l({1\ 0\choose 0\ 0}  M_{k}(0,\sigma)\ri)}\\ 
&=& 
\textstyle {\rm Trace}\l({1\ 0\choose 0\ 0}  
A^{1-\sigma_{k-1}}B^{\sigma_{k-1}} 
M_{k-1}(0,\sigma_1,\ldots,\sigma_{k-2})
\ri)\\
&=& \textstyle
{\rm Trace}\l(\l({1\ 0\choose 0\ 0}+\sigma_{k-1} {0\ 1\choose 0\ 0}\ri) 
 M_{k-1}(0,\sigma_1,\ldots,\sigma_{k-2})\ri)\\
 &=&\textstyle\hcb_{k-2}(\sigma_1,\ldots,\sigma_{k-2})+
\sigma_{k-1}\hcb_{k-2}(1-\sigma_1,\ldots,1-\sigma_{k-2})
=\hcb_{k-1}(\sigma)
\eeqno
and similar for the second identity.
Thus 
\beq
T_k(0,\sigma)=T_k(1,1-\sigma)\geq\hcb_{k-1}(\sigma)\qquad(k\in\bN).
\Leq{lower:bound}
{\bf 3)}
Since the (grand) canonical free energies are given by
\[F^{C/G}_k(\beta)
=-\frac{\ln\l( \sum_{\sigma\in\Gb_{k}}\l({\bf h}^{C/G}_k(\sigma)\ri)^{-\beta} \ri)}
{\beta k} ,\] 
these two inequalities imply
\[\frac{k-1}{k}F^C_k(\beta)-\frac{\ln2}{\beta k}\leq F_k(\beta) \leq 
F^G_k(\beta).\]
The canonical and grand canonical ensembles have the same limit free energy, 
since
\beq
F^C_k \leq F^G_k\leq F^C_k + \frac{\ln(k+2)}{k}.
\Leq{FFF}
So the limit free energy $F$ of the Farey chain coincides with 
the one of the number-theoretical spin chain

The lower inequality in (\ref{FFF}) follows from (\ref{def:hg}), the upper
inequality from \<\hgb_k\leq (k+2)\cdot \hcb_k\>, which is a consequence
of (\ref{def:hg}) and the relation
\[\hcb_k(1-\sigma)\leq (k+1) \cdot\hcb_k(\sigma)\qquad (\sigma\in\Gb_k)\]
(which follows from Def.\ (\ref{def:hck}) by induction).\hfill \< \Box \>\\[2mm]
The same conclusion was reached in \cite{KO} by a different method.
\begin{coro}
The Farey spin chain has exactly one phase transition, at \< \beta=2 \>. 
\end{coro}
{\bf Proof.} 
This follows from the corresponding statement in \cite{CK} 
for the number-theoretical spin chain.\hfill \< \Box \>
\section{Low Temperature Magnetization}
Due to the invariance of the energy function \< E_k \> w.r.t.\ spin flips,
the mean magnetization \< m_k \> has expectation zero.
However, the long-distance correlations are measured by the square of
that variable. 
\begin{theorem} \label{low}
In the low temperature phase \< \beta>2 \>
\[\lim_{k\ar\infty} \LA m_k^2\RA_k(\beta) =1.\]
\end{theorem}
{\bf Proof.} 
We complement estimate (\ref{lower:bound}) by 
\[T_k(\sigma)>k\qquad (\sigma\in\Gb_k,(0,\ldots,0)\neq\sigma\neq
(1,\ldots,1)),\]
which follows inductively from Def.\ (\ref{def:Mk}) by noticing
that for such \< \sigma \> both off-diagonal entries are \< \geq1 \>.
We thus have
\beq
0\leq Z_k(\beta)-2\cdot 2^{-\beta} \leq \sum_{n=1}^\infty a_k(n) n^{-\beta}
\Leq{Zk:in}
with 
\[a_k(n):=|\{ \sigma\in\Gb_k\mid \max(\hcb_k(\sigma),k+1)=n\}|.\]
It is known \cite{Kn1} that 
\[Z_k^c(\beta):= \sum_{\sigma\in\Gb_k} \hcb_k(\sigma)^{-\beta}\]
can be written in the form 
\[Z_k^c(\beta)=\sum_{n=1}^\infty \Euler_k(n) n^{-\beta}\]
with 
 \>\Euler_k(n)\leq \Euler(n) \>
and \< \Euler_k(n) = \Euler(n) \> for \< n\leq k+1 \>,
\[\Euler(n):=
| \l\{ i\in \{1,\ldots,n\} \mid {\rm gcd}(i,n)=1 \ri\} |\] 
being Euler's \< \varphi \>-function.
Thus for \< \Re(\beta)>1 \>
\[\lim_{k\ar\infty} Z_k^c(\beta)=\sum_{n=1}^\infty \Euler(n) n^{-\beta} = 
\frac{\zeta(\beta-1)}{\zeta(\beta)}.\]
Substituting the identity
\[\sum_{n=1}^\infty a_k(n) n^{-\beta}=
Z_k^c(\beta)-\sum_{n=1}^k \Euler(n)\l(n^{-\beta}-(k+1)^{-\beta}\ri) \]
for the r.h.s.\ of (\ref{Zk:in}), we thus get
\beqno
\lim_{k\ar\infty} |Z_k(\beta)-2\cdot 2^{-\beta} | &\leq& 
\lim_{k\ar\infty} |(k+1)^{-\Re(\beta)}|\cdot \sum_{n=1}^k \Euler(n)\\
&\leq& 
\lim_{k\ar\infty} |(k+1)^{-\Re(\beta)}|\cdot \sum_{n=1}^k n =0,
\eeqno
so that 
\< \lim_{k\ar\infty} Z_k(\beta)=2\cdot 2^{-\beta} \> for \< \Re(\beta)>2 \>.
Using \<  0\leq m_k^2\leq 1  \> and 
\<  m_k^2((0,\ldots,0)) = m_k^2((1,\ldots,1)) = 1  \>,
we thus get 
\<  \lim_{k\ar\infty} \LA m_k^2\RA_k(\beta)=1.  \>
\hfill \< \Box \>\\[2mm]
This extends the same conclusion, reached by
a different argument for \<\beta > 3\> \cite{KO}.

For the ferromagnetic spin chain the limit mean magnetization
\<\LA m\RA:=\lim_{k\ar\infty}\LA m_k\RA_k\> equals 1 for the canonical ensemble and \<\beta>2\>, whereas
it vanishes in the high temperature region \cite{CK}.

For the grand canonical ensemble, as for the Farey ensemble, \<\LA m\RA\>
vanishes identically, since the interaction is even. Of course this does not 
say much about the actual structure of the extremal Gibbs states.
\section{High Temperature Demagnetization}
Now we consider the mean magnetization in the high temperature regime
\< \beta<2 \>. To show that the expectation of the square vanishes, we
need a correlation inequality.
So consider for \< n\in\bN \> the configuration 
 \< \tau\in\Gb_{n+2}\> with spins \< \tau_1:=\tau_{n+2}:=0\> ,
 \< \tau_l:=1\mbox{ for }2\leq l\leq n+1\> , and the event 
\[\cE_k^n := \{\sigma\in\Gb_{k+n+2}\mid \sigma_l=\tau_l
\mbox{ for }1\leq l\leq n+2\} \]
of an initial string of \< n \> adjacent  1-spins enclosed by  
0-spins.

Due to the long range character of the interaction, one might think that,
given \<\cE_k^n\>, the ferromagnetic interaction would tend
to align the other spins in the  1-direction (equal to
\<\tau_2=\ldots=\tau_{n+1}\>), at least if  
\< n\>  is large. 

This would mean a {\em negative} conditional expectation 
of \<s_i=(-1)^{\tau_i}\> for \<i\in \{n+3,\ldots,n+k+2\}\>. 
Because of the dominance of the multi-body interactions 
this is, however, not the case. 

In fact, the non-inverted spins 
\< (-1)^{\tau_1}=(-1)^{\tau_{n+2}}=1\> tend to produce an anti-ferromagnetic
effective coupling between the spins
in the regions \<2,\ldots,n+1\> and \<n+3,\ldots,n+k+2 \> they separate:
\begin{propo} \label{propo}
For \< \Lambda \subset \{n+3,\ldots,n+k+2\} \> and \< \beta\geq 0 \>
\beq
\LA \l.s_\Lambda\ri|\cE_k^n\RA_{k+n+2} (\beta)\geq 0\qmbox{with}
s_\Lambda:=\prod_{i\in\Lambda}s_i, 
\Leq{Y}
\< \LA f\l| \cE \ri.\RA_{l}\>  denoting the expectation of  
\< f:\Gb_l\ar\bR\> , 
conditioned by the event  \< \cE\> .
\end{propo}
{\bf Proof.} 
We set 
\[T_k^n:\Gb_k\ar\bN\qmbox{,} T_k^n(\sigma):=T_{k+n+2}(\tau,\sigma)\]
and  \< E_k^n:=\ln(T_k^n)\> .
We first prove 
\beq
(\Fou_k E_k^n)(t) \leq 0\qquad (t\in\Gb_k\setminus\{0\}),
\Leq{neq}
using a polymer technique similar to the one above. 
Namely we redefine the set  \< P_k \<  of polymers by
\[P_k:=\l\{p_m^L,p_m^R\ri\}_{1\leq m\leq k}\dot{\cup} 
\l\{p_{l,r}\ri\}_{1\leq l<r\leq k},\]
and map them to the group elements  
\< \hat{p}_m^L:=\hat{p}_m^R:=\delta_m\in\Gb_k\> 
resp.\  \< \hat{p}_{l,r}:=\delta_l+\delta_r\in\Gb_k\>. Depending upon the
length  \< n\>  of the  1-substring, the polymer activities are given by
$$z(p_m^L):=-\frac{3^{-|{\rm supp}(p_m^L)|}}{2(n+1)}\,,\,
z(p_m^R):=-\frac{3^{-|{\rm supp}(p_m^R)|}}{2(n+1)}\mbox{ and }
z(p_{l,r}):= -3^{-|{\rm supp}(p_{l,r})|},$$
where 
$${\rm supp}(p_m^L):=\{1,\ldots,m\},\,{\rm supp}(p_m^R):=
\{m,\ldots,k\}\mbox{ and } {\rm supp}(p_{l,r}):=\{l,\ldots,r\}.$$
Polymers with intersecting supports and the polymers  \< p_m^L\> ,  \< p_m^R\> 
are mutually incompatible.
We now claim that in analogy with (\ref{poly})
the Fourier transform  \< j_k^n:=\Fou_k T_k^n\>  can be written as
\beq
j_k^n(t) = 2(n+1)
\l(\textstyle \frac{3}{2}\ri)^k \sum_{X\in D_k^{\infty}:\hat{X}=t} z(X) 
\qquad(t\in\Gb_k).
\Leq{jkn}
To show this, we write  \< t=(t_1,\ldots,t_k)\>  uniquely in the form
\[t=(0_{m_1},1,0_{m_2},1,\ldots,0_{m_u})\qquad (m_i\geq0)\]
so that 
\[j_k^n(t) = 2^{-k}{\rm Trace}\l(NS^{m_1}DS^{m_2}D\ldots S^{m_u}\ri)\]
with  \< N:=AB^nA=(n+1)\cdot {1\ 1\choose 1\ 1} + D\> .
\begin{itemize}
\item
If  \< u\>  is odd, then
\[j_k^n(t) = 
(n+1)\cdot (-1)^{(u-3)/2}2^{-k}{\rm Trace}
\l({\textstyle {1\ 1\choose 1\ 1}}S^{\Delta m_1}DS^{\Delta m_2}D\ri) \]
with \< \Delta m_1 := \sum_{i=1}^{(u+1)/2}m_{2i-1} \> and
\< \Delta m_2 := \sum_{i=1}^{(u-1)/2}m_{2i} \>. So
\beqno
j_k^n(t) &=& 
2(n+1)\cdot (-1)^{(u-1)/2}2^{-k}3^{\Delta m_1}\\ 
&=& 2(n+1)\cdot \l(\textstyle \frac{3}{2}\ri)^k 
\prod_{i=1}^{(u-1)/2}(-3^{-m_{2i}-2}).
\eeqno
\item
If  \< u \>  is even, then
\[j_k^n(t) = 
(-1)^{u/2-1}2^{-k}{\rm Trace}
\l(DS^{\Delta m_1}DS^{\Delta m_2} \ri) \]
with \< \Delta m_1 := \sum_{i=1}^{u/2}m_{2i-1} \> and
\< \Delta m_2 := \sum_{i=1}^{u/2}m_{2i} \>. So using (\ref{dsd})
\beqno
j_k^n(t) &=& 
(-1)^{u/2}2^{-k}(3^{\Delta m_1}+3^{\Delta m_2})\\ 
&=& 2(n+1)\cdot \l(\textstyle \frac{3}{2}\ri)^k \cdot
\l[\l(-\frac{3^{-m_{u}-1}}{2(n+1)}\ri)\cdot
\prod_{i=1}^{u/2-1}(-3^{-m_{2i}-2})\ri.+\\
& & \hspace{3cm}\l.\l(-\frac{3^{-m_{1}-1}}{2(n+1)}\ri)\cdot
\prod_{i=1}^{u/2-1}(-3^{-m_{2i+1}-2})\ri].
\eeqno
\end{itemize}
In both cases this coincides with the r.h.s.\ of (\ref{jkn}).
Similar as in Lemma \ref{l2}, we get 
\[
(\Fou_k E_k^n)(t)=
\sum_{\stackrel{X\in {C}^{\infty}_k}{\hat{X}=t}} 
\frac{n(X)}{|X|!} z^X 
\qquad(t\in\Gb_k\setminus\{0\}),
\]
from which (\ref{neq}) follows, using (\ref{sign:n}) and the negativity of all
polymer activities. Now
\[\LA \l.s_\Lambda\ri|\cE_k^n\RA_{k+n+2} (\beta) = \frac{\sum_{\sigma\in\Gb_k}
s_\Lambda(\sigma)e^{-\beta E_k^n(\sigma)}}{\sum_{\sigma\in\Gb_k}
e^{-\beta E_k^n(\sigma)}},\]
so that (\ref{Y}) is a consequence of the first GKS inequality for ferromagnets.
\hfill $\Box$
\begin{theorem} \label{high}
In the high temperature phase \< 0\leq \beta<2 \>
\[\lim_{k\ar\infty} \LA m_k^2\RA_k(\beta) = 0.\]
\end{theorem}
{\bf Proof.} 
Since by translation invariance 
\<\LA m_g^2\RA_g = \frac{1}{g} \sum_{j=1}^g\LA s_1 s_j\RA_g \>, it
suffices to show that for \< \vep>0 \> 
there is a uniform correlation estimate of the form
\beq
|\LA s_1 s_j\RA_g(\beta)| \leq \vep
\qquad(j\in\{j_0(\vep),\ldots,g-j_0(\vep)\}). 
\Leq{vept}
We consider the family 
\< \l\{\cE^{n,l}_{g-n-2} \ri\}_{\stackrel{n=1,\ldots,\nmax}{l=1,\ldots,n}}\>
of events
\[\cE^{n,l}_{g-n-2}:=\cS_k^{-l}\cE^{n}_{g-n-2}\subset
\Gb^1_g\qmbox{with}\Gb^j_g:=\{\sigma\in\Gb_g\mid \sigma_j=1\},\]
using the shift map (\ref{shift}) on \< \Gb_g \>.
As these events are disjoint, 
\beq
\sum_{n,l}\bP_{\beta,g}(\cE^{n,l}_{g-n-2}) =
\bP_{\beta,g}\l(\bigcup_{n,l}\cE^{n,l}_{g-n-2}\ri)\leq
\bP_{\beta,g}(\Gb^1_g) =\eh
\Leq{eh}
for the Gibbsian probability 
\< \bP_{\beta,k}(\cE):=\sum_{\sigma\in\cE}e^{-\beta E_k(\sigma)}/Z_k(\beta) \>
of an event \< \cE \>.
On the other hand if \< \beta<2 \>, for \< \vep>0 \> there is a 
\< \nmax(\vep)\> with 
\beq
\sum_{n=1}^\nmax \sum_{l=1}^n \bP_{\beta,g}(\cE^{n,l}_{g-n-2})\geq 
\eh (1-\vep)
\Leq{ehvep}
for all large \< g \>. This property, which is specific to the high-temperature
region, can be proved as follows.
We note that the thermodynamic limit of the internal energy 
\[U:= \lim_{k\ar\infty} U_k\qmbox{with} 
U_k:= \LA {\textstyle \frac{1}{k}}E_k\RA_k\]
exists and equals \<U(\beta)=\frac{d}{d\beta} \beta F(\beta)\>. 
By concavity and analyticity of \<\beta\mapsto \beta F(\beta)\>, and by
\<F(\beta)=0\> for \< \beta\geq2 \> we conclude that 
\[U(\beta)>0\qquad (\beta<2).\]
This implies a positive limit density of spin flips between neighbouring spins and
thus the existence of an \<\nmax(\vep)\> meeting (\ref{ehvep}).

Since by spin inversion symmetry
\[\LA s_1 s_j\RA_g = -2\l( \LA s_j\,|\,\cC_g \RA_g\cdot \bP_{\beta,g}(\cC_g) +
\sum_{n,l}\LA s_j\,|\, \cE^{n,l}_{g-n-2} \RA_g\cdot \bP_{\beta,g}(\cE^{n,l}_{g-n-2})
\ri)\]
for \< \cC_g := \Gb^1_g\setminus\cup_{n,l} \cE^{n,l}_{g-n-2} \>,
by (\ref{eh}), (\ref{ehvep}) and Proposition \ref{propo}
\[\LA s_1 s_j\RA_g \leq 2\bP_{\beta,g}(\cC_g)\leq \vep.\]
Together with a converse estimate this proves (\ref{vept}).
\hfill $\Box$
\\[1cm]
{\bf Acknowledgement.} One of the authors (P.C.) would like to thank 
the Max-Planck-Institute for Mathematics in the Sciences (Leipzig), 
where this work was done.

\end{document}